# TeleCrowd: A Crowdsourcing Approach to Create Informal to Formal Text Corpora

Vahid Masoumi[a], Mostafa Salehi[a,*], Hadi Veisi[a], Golnoush Haddadian[b], Vahid Ranjbar[c], Mahsa Sahebdel[a]

[a] *Faculty of New Sciences and Technologies, University of Tehran, Tehran, Iran*
[b] *Applied Linguistics, Language and Linguistics Center, Sharif University of technology, Tehran, Iran*
[c] *Department of Computer Engineering, Yazd university, Yazd, Iran*

**Abstract**

Crowdsourcing has been widely used recently as an alternative to traditional annotations that is costly and usually done by experts. However, crowdsourcing tasks are not interesting by themselves, therefore, combining tasks with game will increase both participant's motivation and engagement. In this paper, we have proposed a gamified crowdsourcing platform called TeleCrowd based on Telegram Messenger to use its social power as a base platform and facilitator for accomplishing crowdsourcing projects. Furthermore, to evaluate the performance of the proposed platform, we ran an experimental crowdsourcing project consisting of 500 informal Persian sentences in which participants were supposed to provide candidates that were the formal equivalent of sentences or qualify other candidates by upvoting or downvoting them. In this study, 2700 candidates and 21000 votes were submitted by the participants and a parallel dataset using candidates with the highest points, sum of their upvotes and downvotes, as the best candidates was built. As the evaluation, BLEU score of 0.54 was achieved on the collected dataset which shows that our proposed platform can be used to create large corpora. Also, this platform is highly efficient in terms of time period and cost price in comparison with other related works, because the whole duration of the project was 28 days at a cost of 40 dollars.

[*]Corresponding author

  *Email addresses:* v.masoumi@ut.ac.ir (Vahid Masoumi), mostafa_salehi@ut.ac.ir (Mostafa Salehi), h.veisi@ut.ac.ir (Hadi Veisi), g_haddadian@alum.sharif.ir (Golnoush Haddadian), vranjbar@yazd.ac.ir(Vahid Ranjbar), sahebdel.mahsa@ut.ac.ir (Mahsa Sahebdel)

## 1. Introduction

Most of the tasks in Natural Language Processing area require language corpora. Normalizing informal sentences to their formal equivalents is one of the area of NLP tasks that can be seen as machine translation tasks. However, The quality of machine translation systems that are based on machine learning approaches is strongly related to the number of parallel corpora that is available for the languages. However, most languages such as Persian have limited or no readily available parallel corpora. There are two ways of collecting such corpora. The first way is using linguistic experts to annotate corpora, but it is costly and time-consuming. For example, Bijankhan [1] is the one of the largest PoS tag corpora in Persian that is annotated in around five years using language experts. Second, crowdsourcing (CS) is by far inexpensive and time-saving [2] and it is widely using as a way of collecting large data in studies [3, 4, 5, 6, 7, 8]; However in this regard, the quality control of the annotated labels is such a challenge that should be carefully handled in order to have high quality labels [9]. An active crowd of participants is an indispensable part as the key to any CS project and as a result, the participants' motivation is of great importance and should be taken into great consideration [10]. In this respect, if the tasks are combined with games, the CS participants will find them more interesting and engaging, therefore, as the motivation of work increases, its completion time decreases.

Some of the messengers such as Telegram, a cloud-based instant messaging application, in spite of filtering in Iran, are hugely popular in a way that most percent of the internet traffic in Iran is being spent on this popular messenger and it has more than 40 million active users in this country [11]. Therefore, the great usage of this messenger is an outstanding opportunity to build a CS platform that motivates people to solve Human Intelligence Tasks (HITs). In this paper, we proposed TeleCrowd, a gamified crowdsourcing platform, based on Telegram Messenger to use its power as a social media to build a CS platform. Moreover, this platform is able to handle other kinds of HITs such as image annotation and audio transcription as well. The goal of this corpus collection is to augment corpus resources for NLP investigations. As the case study, a parallel dataset of informal Persian sentences with their formal equivalents was collected via the proposed platform to show its efficiency.

The rest of the paper is organized as follows. First, a brief review of the relevant studies are presented in Section 2 and then in Section 3, our proposed CS framework is described. Further in Section 4, our results and evaluation metrics used to measure the accuracy of the corpus is discussed and finally, Section 5 discusses future work and conclusion.

## 2. Background: Crowdsourcing Based Corpus Creation

Crowdsourcing means outsourcing works, tasks, and problem-solving to online people rather than to employees or other offline people [2]. It has been considered to be a precious way to solve problems which are arduous for computers to solve but are easy for humans and therefore, they are outsourced to humans. Crowdsourcing can be categorized into four major categories [12]. The First category is crowdsolving which use each individuals contribution of the crowds to solve heterogeneous problems. This approach is often used for complex problems that usually have no pre-definable solutions. The second category is crowdcreating that help to create comprehensive artefacts contributed by heterogeneous solutions. The third category deals with crowdrating approaches which use the wisdom of crowds for homogenous problems and they are commonly used for prediction or validation. And finally, the fourth category is crowdprocessing which rely on the crowd to perform large quantities of homogenous tasks.

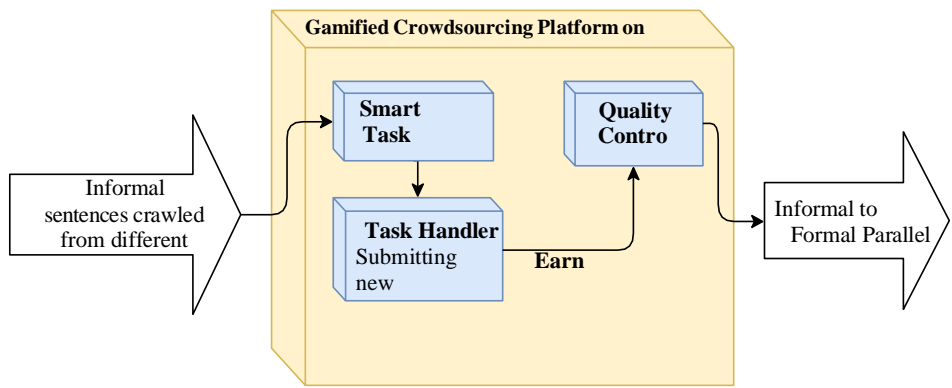

Figure 1: Architecture of proposed framework

Most studies in the area of collecting annotated corpora have used available crowdsourcing platforms such as Mturk [13, 7, 4] that take place in crowdprocessing category. In [13], with the leverage of crowdsourcing through Mturk, a list of labeled Out-Of-Vocabulary (OOV) words is generated from a 180-million-word twitter corpus. Since crowdsourcing is really helpful for low resource languages, Mturk as an efficient crowdsourcing platform is used to create a collection of parallel corpora between English and six languages from Indian subcontinent naming Bengali, Hindi, Malayalam, Tamil, Telugu, and Urdu [4]. Moreover, in [7] a system is developed to extract prepositional phrases and their potential attachments from ungrammatical and informal sentences and posed subsequent disambiguation tasks in form of multiple choice questions to participants. Collecting highly parallel data in Mturk may be expensive if tasks are not designed properly so in [5] a novel data collection framework is proposed in which annotators had to watch a very short video clip for each task and describe the main action or event of the video clip in one sentence, in the language of their choice.

Although Mturk brings some advantages for collecting annotated corpora, it is not helpful in all kinds of purposes. Therefore, the importance of developing a crowdsourcing platform for different goals has been considered in studies [6, 8]. [6] explored the use of Wechat Official Account Platform (WOAP) in order to build a speech corpus and aimed to assess the feasibility of using WOAP followers (also known as contributors) to assemble speech corpus of Mongolian. [8] provided a user-friendly online interface for crowdsourced annotation tasks and their aim was to create a publicly available Russian paraphrase corpus which could be applied for information extraction, text summarization, and compression.

Therefore, in this study, a crowdsourcing platform based on Telegram Messenger is proposed to tackle this problem and additionally, this platform can be utilized for any other purposes such as image annotation, audio transcription and other types of data labeling in everywhere.

**3. Proposed Crowdsourcing Framework**

Since Telegram is a popular messenger application in Iran, there is such a great opportunity to use this popularity and use its social power to accomplish crowdsourcing projects such as creating large text corpora. Users can accomplish HITs in their spare time while they are using their messengers everywhere and get paid. Besides, as far as we know, there is not such an informal to formal parallel Persian corpus, collected by crowdsourcing approaches, and all available corpora are manually created by language experts. Therefore, in the experimental part, a parallel corpus of informal to formal Persian text is created to show its performance and efficiency.

The overview of TeleCrowd is demonstrated in fig. 1 which consists of different parts which are described briefly in the following. First, texts were crawled from different sources such as Telegram, Twitter and Digikala's Product Reviews[1] and 500 sentences with informal forms were selected manually. The yellow box shows the Telegram Bot which is combined with gamification and each one of the gamification elements which have been used in TeleCrowd are described in Section 3.2.

The proposed platform has three main submodules: *Smart Task Dispatcher*, *Task Handler* and *Quality Control*. *Smart Task Dispatcher* is responsible for assigning tasks to participants in a way that all tasks have a uniform number of labels. After that, *Task Handler* module is responsible to handle tasks which are performed by the participants. In this study, tasks are informal Persian sentences in which participants are asked to either provide a new candidate for them or upvote/downvote on a candidate that is already submitted by other participants. The last submodule of the platform is *Quality Control* which is extremely important since text normalization tasks are not straightforward for most of the participants. In this regard, a collaborative technique has been proposed which helps participants to learn from each other and improve others' contribution. Also, as a part of the quality control system, gamification elements are considered as well to improve workers' performance and overall quality. Quality control technique will be described in more details in 3.3.

*3.1. Communication with Telegram servers*

Fig. 2 shows the connections between us and the Telegram server which is based on HTTPS protocol. As this figure illustrates, when a user makes an action like submitting a new candidate or vote, a request containing callback data is sent to us by Telegram, and then, an appropriate response is sent back to Telegram based on the user action.

---
[1]Digikala.com is the biggest E-Commerce in the Middle East

Finally, the provided response appears on the user's Telegram client with a proper appearance.

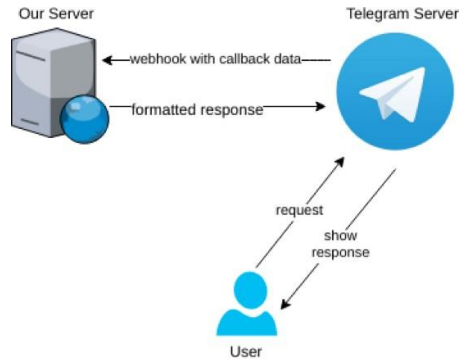

Figure 2: Communications with telegram servers

*3.2. Gamifying the Framework*

Gamification refers to a design that attempts to firstly, increase the intrinsic motivation of participants to engage in a given activity and secondly, to increase or otherwise change the given behavior. Besides, other studies in the gamification area show that combining game with crowdsourcing will increase engagement of the users, improve the quality of the users' answers and also, reduce the total cost of crowdsourcing by expediting the labeling task process which results in reducing the total required time of crowdsourcing project [10]. If we consider gamification in the context of crowdsourcing, gamification can be seen as an attempt to redirect participants' motivation from purely rational gain-seeking to self-purposeful intrinsically motivated activity. Gamification elements have been widely used in various crowdsourcing investigations. Such elements include Point, Leaderboard, Achievement, Level, Progress, Feedback, Reward, Storytelling, Missions, and Virtual Territories. Since our crowdsourcing project fall into *crowdrating* and *crowdprocessing* categories, the most relevant elements of gamification related to these categories are Point, Leaderboard, Achievement, Level, and Progress according to [10].

- Point: each candidate can be upvoted or downvoted and the candidate's point is the sum of its up and down votes as seen in 1. For example, when a candidate has 3 upvotes and 1 downvote, the point of the candidate will be 2.

$$Point(candidate) = Upvotes + Downvotes \quad (1)$$

- Leaderboard: participants can see their ranks on the leaderboard and it will motivate them to be on the top of the leaderboard to win the final prize which is described in section 3.4. Also, a reminder was sent to the participants each day in order to increase the retention rate by informing them of their ranks and their difference with the next competitor in the leaderboard to motivate them to earn more points.

- Achievement: to increase the participants' motivation and improve the quality of the answers, some achievements were designed. For instance, participants tend to submit upvotes rather than downvotes and since downvotes are helpful to detect wrong answers and also help other participants to avoid mistakes, an achievement is designed to encourage participants to detect the low-quality answers by submitting downvotes on them.

- level: logarithmic equation (2) is designed in a way that participants can achieve levels easily in the early stages of the game but as their points increase, reaching to the higher levels of the game will be harder.

$$Level(x) = 0.4 \times (2x + 130 - 20) \qquad (2)$$

- Progress: indicates what percentage of the tasks is done by the participant. Thus, it helps participants to estimate the total amount of required time to complete all the remaining tasks.

*3.3. Quality Control*

Some crowdsourcing tasks such as text normalization are complicated for people. Hence, a *collaborative* technique was employed to achieve high-quality results and also, accelerate the process of task completion for even those with low skills. In the proposed technique, participants could submit new candidates with the help of other participants' candidates or vote on candidates based on other participants' votes. All the candidates in the list were sorted descendingly based on their points, therefore, the best candidates with the highest point were always on the top of the list and the candidates with a specific number of negative points were disappeared. At the end of crowdsourcing, the best candidates that had the highest points, could be extracted as the qualified labels.

Moreover, *crowdprocessing* and *crowdrating* projects are run in two sequential phases. First, a *crowdprocessing* project is created and participants annotate tasks and after that, in a *crowdrating* project participants who had high quality in annotation step will be permitted to rate the answers as the validation phase and they will earn more money compared to the labeling step since they are qualified; however, this approach is costly and also time-consuming. Therefore, to decrease the overall time for project completion, *crowdprocessing* and *crowdrating* steps were combined and participants were allowed to submit a candidate or vote on a candidate simultaneously, but it should be noted that sentences without any candidates were sent first and so, there was always a candidate to vote on.

Besides, to measure the quality of the participants' work, Expectation Maximization (EM) technique [14] is utilized which transforms the aggregation problem into a maximum likelihood formulation, in which human inputs are sample values and the aggregated results are parameters to be estimated. The EM technique iteratively computes object probabilities in two steps: Expectation (E) and Maximization (M). In the (E) step, object probabilities are estimated by weighting the answer of participants according to the current estimates of their expertise. In the (M) step, the participants' expertise is re-estimated based on the current probability of each object. This iteration goes on until all object probabilities are unchanged.

*3.4. Incentive*

There are many ways to motivate participants in crowdsourcing projects such as awarding the winner a prize or a little amount of money for accomplishing each task. Since the number of tasks in this project was not much enough to reward the participants per task and also, they could not earn a significant financial reward, only the top participant with the highest point on the leaderboard was awarded financial reward of 40 USD. So in this way, competition between participants would be much more intense and they would do their the best to be the top one on the leaderboard.

*3.5. System Flow*

In this section, we are going to describe the flow of the system from the first moment that a participant interacts with the system. First, the variables that were used as thresholds for each part of the system are described in table 1.

Table 1: Variables

| Variable | Description | Threshold |
|---|---|---|
| $\alpha$ | Total allowed number of candidates in the candidate list | 4 |
| $\beta$ | Threshold for each candidate in order to be stayed in the list | -3 |
| $\delta$ | Threshold for the agreement on a low quality candidate | -3 |
| $\eta$ | Threshold for the agreement on a high quality candidate | 10 |
| $\theta$ | Points that participants will earn for the first time by submitting new candidate/vote | 10 |
| $\mu$ | Points that participants will earn for the second time by submitting new candidate/vote | 5 |
| $\gamma$ | Maximum number of candidate/vote submission for each user on each candidate | 2 |
| $\rho$ | Amount of points that participants will be charged if they get down vote by other users | -5 |

The flow of the system is as follows: when a participant submits a new answer, first, it checks whether the answer is a vote or a candidate. If the total number of available candidates in the list is less than $\alpha$, then the participant can not add a new candidate until one of the candidates is eliminated from the list by downvotes of other participants and its point gets lower than $\beta$, so the participant should vote one of the candidates or skip the task and go to the next one. For increasing agreement between the participants, if the number of submitted downvotes on a candidate is higher than $\delta$, then all the participants who agreed on low quality candidate, will achieve $\mu$ points per voter and the participant who submitted the candidate will be punished by $\rho \times$ *No. of voters who downvoted the candidate*. This is also true for upvote but with the threshold of $\eta$ since users tend to submit upvote rather than downvote. Moreover, $\theta$ and $\mu$ are respectively the points that participants will earn by submitting a new candidate for the first and the second time.

## 4. Results and Evaluations

In this section, we are going to discuss our dataset which was crawled from different sources. Evaluation metrics for measuring corpus accuracy will be discussed in Section 4.2. In the next parts, collaboration statistics of participants with TeleCrowd and the accuracy of the collected dataset through our platform will be described.

### 4.1. Dataset

Since the experimental project which was selected to run on our proposed crowdsourcing platform was informal to formal text corpus creation, several scrapers were designed in order to crawl different sources including Twitter, Telegram and Digikala's product reviews. We crawled Telegram for 6 months and around 1.3M messages of different channels and groups were gathered. The source code of our Telegram Crawler is available on github[2]. Also, we crawled Twitter for 3 months and around 1.3M tweets were gathered as well. Although tweets and Telegram posts were sufficient enough, in order to have more diverse sentences, 12K Digikala's product reviews are added to increase the users' engagement.

### 4.2. Evaluation Metrics

BLEU [15] is one the most popular metrics for evaluating machine translation accuracy in linguistic studies between a source sentence in one language and target sentences in another language. Regarding the fact that converting informal sentences to their standard forms problem was mapped to the machine translation problem with this difference that here, source sentences were informal sentences and the target language were their standard equivalents. Therefore, BLEU was used as the evaluation metric to indicate how much the provided equivalents by the participants are close to the experts' equivalents. The BLEU score is a number in the range of [0,1] which higher values close to 1 show that translated sentences are close to human translations whereas, the BLEU scores close to zero show low-quality answers with high certainty. Answers with the BLEU scores higher than 0.5 have acceptable quality. Also, we gathered 846 sentences using 4 experts as our reference set for BLEU metric which was almost 1.69 references per candidate. It is not necessary to have more than 3 references per candidate since BLEU in corpus level compares each candidate with all the references in the corpus.

---

[2]https://github.com/vhdmsm/Telegram-Crawler

*4.3. Collaboration Statistics*

In this section, demography of the participants and their collaborations with TeleCrowd is discussed. As fig. 3 illustrates, most of the participants are in the range of 20-30 years old and they are distributed equally between men and women.

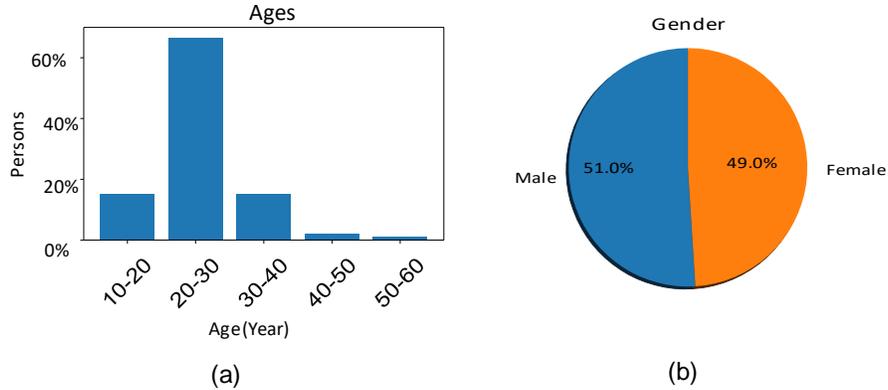

(a)      (b)

Figure 3: participants' Demography

Summary of statistics about the submitted candidates and votes are shown in table 2. Accordingly, the number of submitted votes are much greater than the submitted candidates since submitting a new candidate needs more skills and is much harder than voting on a candidate and besides, each of the informal sentences on average, does not have more than 5 standard forms.

Table 2: Statistics of collected dataset

| Type | Count |
| --- | --- |
| Total submitted candidates | 2700 |
| Total submitted votes | 21000 |
| Average number of candidates per informal sentence | 5.42 |
| Average number of votes per informal sentence | 42.04 |
| Average number of upvotes per informal sentence | 28.61 |
| Average number of upvotes per informal sentence | 13.44 |

Fig. 4 shows the number of candidates for each sentence and as it can be seen, above 50 percent of the sentences have 4-6 candidates.

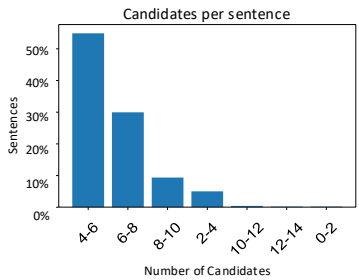

Figure 4: Number of candidates per sentence

As fig. 5 illustrates, the number of agreeing votes are higher than opposing votes. It can be inferred that users tend to leave an agreeing vote rather than opposing votes. For example, around 40 percent of the sentences have between 50 to 60 agreeing votes and around 32 percent of the sentences have between 10 to 30 opposing votes.

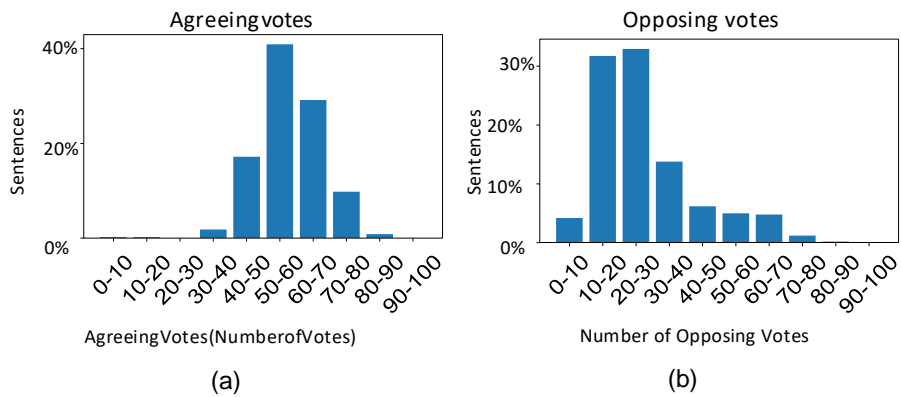

Figure 5: Percentage of agreeing and opposing votes

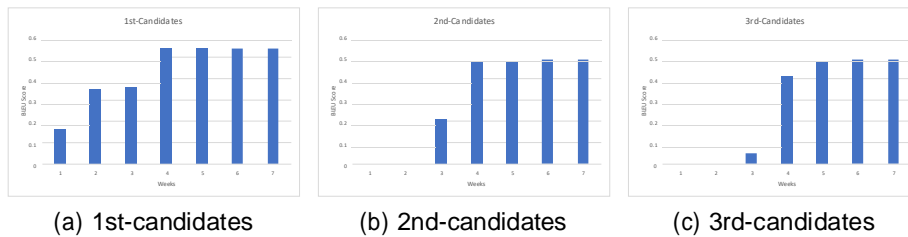

(a) 1st-candidates  (b) 2nd-candidates  (c) 3rd-candidates

Figure 6: BLEU scores over weeks

Submitting a candidate takes more time as it is obvious but voting on a candidate takes less time so users tend to vote because it is easier and faster. We can see this fact in fig. 7.

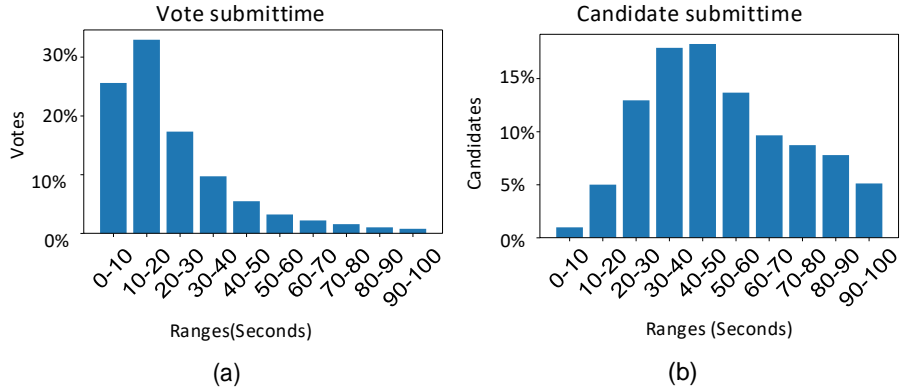

Figure 7: Candidates and votes submit time

*4.4. Evaluation of the Collected Dataset*

As it was described in Section 4.2, BLEU was used for evaluating the collected dataset. Fig. 8 shows the accuracy of BLEU over best nth-candidates in which accuracy of 0.54 was achieved over 1st-candidates and around 0.49 for both of the 2nd and 3rd candidates. The accuracy that is achieved on BLEU score shows that normalized sentences provided by the participants were close to experts' equivalents which were resulted in a high-quality dataset.

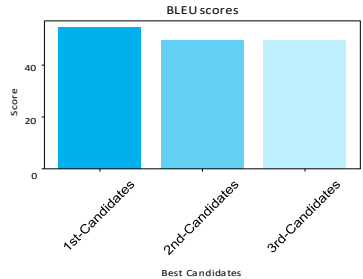

Figure 8: BLEU scores over best candidates

Additionally, fig. 6 shows BLEU scores over weeks from the start of the project and as it can be seen, the best result of BLEU was achieved in week 4, therefore, we could terminate the project at that time since BLEU score had no improvement after that.

As another evaluation metric, the quality of the participants' work was measured through EM technique which was described in the previous sections. Accordingly, the average estimated quality of the participants is demonstrated in fig. 9 in which x-axis is the number of participants who provided n-number of labels and y-axis is the average of estimated quality for each part. Moreover, it can be seen that participants with the higher number of labels were more experienced as they answered more questions, therefore, they could provide better answers compared to the participants with the lower number of answers.

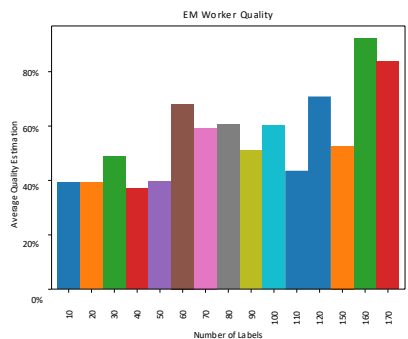

Figure 9: Quality of the participants

## 5. Conclusion and Future Work

Crowdsourcing is a low-cost alternative for traditional labeling methods in which experts label all the corpus manually. Since most users spend a lot of time in messengers like Telegram, such messengers can be considered as a base platform to build a crowdsourcing platform on based on them. Moreover, bringing these users to accomplish crowdsourcing tasks in return of a certain amount of money is not hard, therefore, we benefited from Telegram and its Bot system to propose a crowdsourcing platform called TeleCrowd and as an experimental project, an informal to formal dataset including 500 informal Persian sentences with their standard equivalent was collected using our platform. As a result of the experiment, 2700 candidates and 21000 votes were submitted and by extracting candidates with the highest points as the equivalents, BLEU score of 0.54 was achieved which shows that our proposed platform can be used to create large corpora. Also, from the cost and time perspective, the project was completed in 28 days with a cost of 40 USD which in comparison with related works, it is both time and cost saving. As the future work, we mainly focus on expanding our dataset scale to be much more larger and then as the application of our corpus, we plan to train a deep neural network such as OpenNMT [16] over our corpus to build an autonomous text normalizer which is able to convert informal texts to their standard forms.